# Novel magnetoelectric effects via penta-linear interactions

Hong Jian Zhao[1,2], M. N. Grisolia[3], Yurong Yang[2], Jorge Íñiguez[4,5], M. Bibes[3], Xiang Ming Chen[1,*], and L. Bellaiche[2,*]

[1]*Laboratory of Dielectric Materials, School of Materials Science and Engineering, Zhejiang University, Hangzhou 310027, China.*
[2]*Physics Department and Institute for Nanoscience and Engineering, University of Arkansas, Fayetteville, Arkansas 72701, USA.*
[3]*Unité Mixte de Physique CNRS/Thales, 1 Av. A. Fresnel, Campus de l'Ecole Polytechnique, 91767 Palaiseau and Université Paris-Sud, 91405 Orsay, France.*
[4]*Materials Research and Technology Department, Luxembourg Institute of Science and Technology, 5 avenue des Hauts-Fourneaux, L-4362 Esch/Alzette, Luxembourg.*
[5]*Institut de Ciència de Materials de Barcelona (ICMAB-CSIC), Campus UAB, 08193 Bellaterra, Spain.*
*email: xmchen59@zju.edu.cn and laurent@uark.edu

Magnetoelectric multiferroic materials, particularly with the perovskite structure, are receiving a lot of attention because of their inherent coupling between electrical polarization and magnetic ordering. However, very few types of direct coupling between polarization and magnetization are known, and it is unclear whether they can be useful to the design of novel spintronic devices exploiting the control of magnetization by electric fields. For instance, the typical bi-quadratic coupling only allows to change the magnitude of the magnetization by an electric field, but it does not permit an electric-field-induced switching of the magnetization. Similarly, the so-called Lifshitz invariants allow an electric-field control of complicated magnetic orderings, but not of the magnetization. Here, we report the discovery of novel direct couplings between polarization and magnetization in epitaxial perovskite films, via the use of first-principles methods and the development of an original Landau-type phenomenological theory. Our results feature *penta-linear* interactions involving the ferromagnetic and anti-ferromagnetic vectors as well as the polar distortions and oxygen octahedral tilting, and permit a number of striking effects. Examples include a continuous electric-field control of the magnetization magnitude and sign, and the discrete switching of the magnetization magnitude. Thus, the high-order, penta-linear couplings demonstrated in this work may open new paths towards novel magneto-electric effects, as well as, spintronic and magnonic devices.

# I. Introduction

Discovering multiferroic materials possessing large or novel couplings between their electric and magnetic orders is an important current research direction. It has the potential to deepen the fundamental knowledge of condensed matter physics, and to result in new technological applications in, e.g., the field of spintronics. Three main, general types of magneto-electric (ME) coupling have been revealed so far in multiferroic perovskites. The first one is the ``traditional'' direct *bi-quadratic* coupling between the magnetization, **M**, and the electrical polarization, **P**, and its associated energy is of the form $\Delta E \sim P^2 M^2$.[1] As a result of such a coupling, application of an electric field modifies the magnitude of the magnetization via the electric-field-induced linear change in the polarization. The second kind of ME coupling gathers the terms that are *linear* in polarization and involves two different magnetic quantities. Examples include (1) the Lifshitz invariants $\Delta E \sim$ **P**·[**L**($\nabla$·**L**)+**L**×($\nabla$×**L**)] and $\Delta E \sim$ **P**·[**M**($\nabla$·**M**)−(**M**·$\nabla$)**M**)] [2-3] where **L** is an antiferromagnetic vector, and (2) the spin-current model [4-6] for which $\Delta E \sim$ (**P**×**e**$_{ij}$)·(**m**$_i$×**m**$_j$), where **e**$_{ij}$ is the unit vector joining the magnetic cations at sites *i* and *j* whose magnetic moments are given by **m**$_i$ and **m**$_j$, respectively. This second type is responsible for non-trivial effects, such as the electric-field-driven change of the propagation direction [7,8] and reversal of the chirality [9] of magnetic cycloids. [Note that couplings being linear in P and potentially permitting magnetization switching can occur in other less-studied structures, such as ilmenites [10].] The third kind of ME effects is based on *indirect* couplings of magnetization and polarization. An illustrative example recently reported relies on the existence of two different *trilinear* terms: a first energy, $\Delta E \sim PQ_1Q_2$, couples the electrical polarization with two other *structural* degrees of freedom, $Q_1$ and $Q_2$ (which can, e.g., be two oxygen octahedral tiltings [11-14], or one octahedral tilting and an antiferroelectric mode [15]); and a second trilinear energy, $\Delta E \sim Q_1\, m_P\, m_S$, that couples one of the two other structural degrees of freedom with the predominant magnetic order parameter, $m_P$, as well as with a secondary magnetic order parameter, $m_s$ [16]. The existence of the first trilinear energy implies that the switching of the polarization reverses $Q_1$ (which is assumed to be *softer* than $Q_2$), which in turn reverses $m_s$, because of the second trilinear term [15]. This third type of coupling between polarization and a magnetic quantity is therefore of *indirect* nature.

Today, after two decades of intense work, the possibilities for direct ME couplings seem exhausted, and most of the research on novel ME effects tends to focus on *indirect* effects as those described above [17]. Indeed, it seems hard to believe that new energies can couple *directly* polarization and magnetization to yield useful effects.

Here, we report results of first-principles calculations that show that such novel couplings, in fact, exist. More precisely, our calculations and analysis of $CaMnO_3$ thin films reveal high-order (penta-linear) interactions that directly couple polarization and magnetization, as well as the antiferromagnetic vector and oxygen octahedral tiltings that characterize the investigated structures. Further, we predict that such couplings permit electric-field-control of the magnitude and sign of the magnetization, by inducing either discrete or continuous changes, providing a new and exciting playground for ME effects.

This article is organized as follows. Section II provides details about the first-principles method used here. Section III reports and describes the first-principles results for $CaMnO_3$ thin films, along with the development of a novel Landau-type phenomenological model to analyze and understand them. Finally, a perspective about the applicability of the present results to other systems, as well as about other related phenomena, is given in Section IV.

**II. Method**

First-principles calculations are performed using the VASP code [18] within the framework of local spin density approximations (LSDA) [19]. Note that, as in Ref. [20], we introduce no Hubbard correction for the $Mn^{4+}$ ion. The projector augmented wave (PAW) scheme [18] is employed in the present calculations, with the following electrons explicitly considered: calcium's $3p^6$ and $4s^2$, manganese's $3d^6$ and $4s^1$, and oxygen's $2s^2 2p^4$. The energy cutoff is chosen to be 500 eV. We presently consider four different phases of epitaxial CaMnO$_3$ films, namely the non-polar, so-called c-ePbnm and ab-ePbnm phases [21,22], and the polar Pmn2$_1$ and Pmc2$_1$ states. The lattice vectors of the c-ePbnm, Pmn2$_1$ and Pmc2$_1$ phases are given by $\mathbf{a}=a_{IP}(\mathbf{x}+\mathbf{y})$, $\mathbf{b}=a_{IP}(-\mathbf{x}+\mathbf{y})$ and $\mathbf{c}=(2a_{IP}+\delta)\mathbf{z}$, where $\mathbf{x}$, $\mathbf{y}$ and $\mathbf{z}$ are the unit vectors along the pseudo-cubic [100], [010] and [001] directions, respectively, and where δ is a coefficient to be relaxed while $a_{IP}$ is the in-plane lattice constant of the substrate (the pseudo-cubic setting is used throughout in this paper). Figure 1 schematizes the epitaxial growth of the c-ePbnm, Pmn2$_1$ and Pmc2$_1$ states on a cubic substrate. Regarding ab-ePbnm, its lattice vectors are chosen to be $\mathbf{a'}=2a_{IP}\mathbf{x}$, $\mathbf{b'}=2a_{IP}\mathbf{y}$ and $\mathbf{c'}=\delta'\mathbf{x}+\delta''\mathbf{y}+2a_{IP}\mathbf{z}$, where δ' and **δ"** are parameters to be optimized via structural relaxation, and its growth on a cubic substrate is also depicted in Fig. 1. A k-point mesh of 6×6×4 is used for c-ePbnm, Pmn2$_1$ and Pmc2$_1$ structures, while the selected k-point mesh is 4×4×4 for ab-ePbnm (note its larger unit cell). Note that all four phases considered exhibit an $a^-a^-c^+$ pattern (in Glazer's notation [23]) for the O$_6$ octahedral tilting, where the plus and minus super-scripts indicate anti-phase and in-phase rotations, respectively. For the c-ePbnm, Pmn2$_1$ and Pmc2$_1$ states, the overall anti-phase tilting occurs about the **b** axis, while the in-phase tilting is about the c-axis. Regarding the ab-ePbnm state, the anti-phase tilting is about an axis being close to **a'**+**c'** and the in-phase tilting is about **a'**. Note also that the electrical polarization in the Pmc2$_1$ state lies along the pseudo-cubic [-110] direction (i.e., along the **b** axis), that is, it is parallel to the axis of the anti-phase tilting. In contrast, the electrical polarization in the Pmn2$_1$ phase is oriented along the perpendicular pseudo-cubic [110] direction.

For our structural relaxations, we keep the in-plane lattice parameter ($a_{IP}$) fixed and optimize all the other degree of freedoms, i.e., the aforementioned δ, δ', and δ" parameters and the ionic positions (until the force on each ion is converged within 0.005 eV/Å). The space groups of the resulting relaxed states are determined by using the FINDSYM software [24], and the electric polarization is calculated by the Berry phase method [25].

Regarding magnetic properties, we choose the predominant magnetic ordering to be G-type antiferromagnetic, as in Ref. [20]. In other words, the spin vectors of nearest-neighboring $Mn^{4+}$ ions are anti-parallel to each other, as consistent with the known magnetic configuration of bulk $CaMnO_3$. [26] Spin-orbital coupling is also included when determining the non-collinear magnetic structure. As shown in Fig. 2, we consider two different cases: Case (1) for which the predominant G-type antiferromagnetic vector, **G**, is lying along the pseudo-cubic [001] direction; and Case (2) in which **G** is oriented along the pseudo-cubic [110] direction. The relaxed spin configurations in Cases (1) and (2) are, in fact, the so-called $\Gamma_2(F_a, C_b, G_c)$ and $\Gamma_4(G_a, A_b, F_c)$ magnetic states [27]. The $\Gamma_2$ configuration possesses a weak magnetization along the pseudo-cubic [110] direction (i.e., along the ***a*** vector defining the orthorhombic cell) and a weak C-type antiferromagnetic vector lying along the pseudo-cubic [-110] axis (i.e., along the ***b*** cell vector). On the other hand, for the $\Gamma_4$ magnetic state, the weak magnetization is directed parallel or antiparallel to the pseudo-cubic [001] (***c*** cell vector), and the other (weak) antiferromagnetic vector is now of A-type and is along the pseudo-cubic [-110] (***b*** cell vector). These additional weak magnetic vectors and their directions can be naturally explained by the universal law given in Ref. [16], which couples dominant and weak magnetic orderings with the in-phase and anti-phase tiltings of oxygen octahedra.

Note that, while (as indicated above) we used a plain LSDA functional to treat $CaMnO_3$, we also checked the effect that "Hubbard U" corrections have in the key results of this work. Most importantly, the bare LSDA calculations yield magnetizations that are considerably smaller (0.05 $\mu_B$/f.u. and -0.0498 $\mu_B$/f.u. for the $\Gamma_2$ and $\Gamma_4$ spin configuration, respectively) than those obtained from a LSDA+U calculation with an effective Hubbard $U_{eff}$ of 3.0 eV (0.121 $\mu_B$/f.u. and -0.118 $\mu_B$/f.u., respectively) for bulk $CaMnO_3$. Hence, the effects we discuss are predicted to be quantitatively stronger when a Hubbard correction is included in the simulations. Let us also note that our work has similarities with the investigation reported in Ref. [20], frequently cited in this paper. However, here we go one step further and discuss the two aforementioned (as opposed to only one in Ref. [20]) polar structures that $CaMnO_3$ is predicted to present under tensile epitaxial strain, such a multi-stability being the key to the remarkable magnetoelectric effects we have discovered.

**III. Results**

**A) First-principles results**

Let us first check the accuracy of our simulations to predict structural properties. For that, we perform energy optimization of the $CaMnO_3$ *bulk* in its Pbnm ground state, and find that our a, b, and c equilibrium (orthorhombic) lattice parameters are equal to 5.141, 5.200, and 7.292 Å, respectively. These computed equilibrium lattice parameters are in excellent agreement with the first-principles predictions of Ref. [20], which reports 5.161, 5.205, and 7.309 Å, respectively. On the other hand, as typical of LSDA calculations, they underestimate the corresponding experimental data of Ref. [26] (5.264, 5.278 and 7.455 Å), by about 2.3%, 1.5%, and 2.2%, respectively.

As in Ref. [20], we define the misfit strain experienced by the mimicked $CaMnO_3$ films as $\eta=(a_{IP}-a_0)/a_0$, where $a_0$ is the average value of the a and b lattice parameters (divided by $\sqrt{2}$) of the simulated ground state of bulk $CaMnO_3$. Figure 3 reports the total energy of the studied four phases as a function of $\eta$, in case of tensile strain (that is, for positive $\eta$) when adopting collinear magnetism. The ab-ePbnm phase, which strictly speaking has monoclinic $P2_1/m$ symmetry, is the ground state for (weak) misfit strains having a magnitude smaller or equal to ~ 0.5%. Then, c-Pbnm is the most stable phase for $\eta$ between ~ 0.5% and ~ 3.9%. For higher strains, as emphasized by the inset of Fig. 3, our calculations predict that the $CaMnO_3$ films become polar, the ground state adopting the $Pmn2_1$ space group. Further, a second polar structure with $Pmc2_1$ symmetry is stabilized as well. On one hand, this finding is consistent with the paraelectric-to-ferroelectric transition reported in Ref. [20], for which the critical misfit strain is 3.2%. (The precise value of this critical strain depends on technical details of density-functional calculations [20] and should likely decrease as the temperature increases – as a result of the typical temperature-induced reduction of the electrical polarization.) On the other hand, the predicted polar ground state in Ref. [20] is reported to have $Pmc2_1$ symmetry, while our analysis rather yields $Pmn2_1$. Interestingly, in an investigation of the related compound $CaTiO_3$, Ecklund *et al.* [21] found not only the same two polar structures ($Pmn2_1$ and $Pmc2_1$) but also the same energy hierarchy between them under tensile strain that we find here for $CaMnO_3$, supporting the correctness of our results and analysis.

Interestingly, Fig. 3 also shows that the difference in energies between $Pmn2_1$ and $Pmc2_1$ is enhanced as the strain increases above its critical value of 3.9%. However, this latter energy difference is only of the order of 0.01 eV/f.u. for a strain as large as 7%, and both polar states $Pmn2_1$ and $Pmc2_1$ are more stable than the non-polar ab-ePbnm and c-Pbnm phases for tensile strains larger than 3.9%. It is therefore reasonable to wonder whether there is an "easy" structural path allowing the $Pmn2_1$ phase to transform into the $Pmc2_1$ state, and *vice-versa*, by, e.g., the application of electric fields. In order to resolve such issue, we first select a specific misfit strain, namely 5.8%, and compute the energy *vs* polarization curves for the $Pmn2_1$ and $Pmc2_1$ phases at this strain value. (For any η between 3.9% and 7% the results are qualitatively the same.) These curves are displayed in Fig. 4, where we take the c-ePbnm state as the reference phase for determining the polarization. They indicate that the equilibrium $Pmn2_1$ ground state has an electric polarization of about 29.3 μC/cm$^2$ oriented along the pseudo-cubic [110] direction, while the metastable $Pmc2_1$ state exhibits an electrical polarization of about 24.8 μC/cm$^2$ lying along the perpendicular [-110] direction. The energy along the bridging path between the $Pmn2_1$ and $Pmc2_1$ minima is shown in Fig. 4 as well. These bridging structures therefore possess a polarization having non-zero components along both the pseudo-cubic [110] and [-110] axes, and thus adopt the monoclinic Pm space group. Figure 4 reveals that an energy barrier of 6.5 meV/f.u. (respectively, 0.5 meV/f.u.) needs to be overcome to transit from the Pmn2$_1$ ground state to the stable Pmc2$_1$ phase (respectively, when going from Pmc2$_1$ to Pmn2$_1$). These energy barriers are therefore rather small. As a result, applying realistic electric fields along the pseudo-cubic [-110] direction should allow to switch from the Pmn2$_1$ ground state to the metastable Pmc2$_1$ state, and going back to Pmn2$_1$ from Pmc2$_1$ should be feasible via the application of moderate fields along the pseudo-cubic [110] axis. (Note that the application of in-plane electric fields has been demonstrated in, e.g., Ref. [28].) We will come back to, and take advantage of, this possibility of easily switching back and forth between Pmn2$_1$ and Pmc2$_1$ later on.

Figure 5a shows the weak magnetization of the CaMnO$_3$ films, at a tensile misfit strain of 5.8%, when varying the magnitude and sign of the electrical polarization in the Pmn2$_1$ and Pmc2$_1$ states (note that such variations can be practically done by applying electric fields). One can see that, within the range of investigated polarizations, the weak magnetization associated with $\Gamma_2$ or $\Gamma_4$ reaches its *maximum* value when the polarization vanishes -- that is, when the corresponding structural phase is, in fact, the paraelectric c-ePbnm. Moreover, an interesting magneto-electric effect develops for the four different cases (Pmn2$_1$ and Pmc2$_1$ states with either $\Gamma_2$ or $\Gamma_4$ spin structure), when the polarization is switched on and varies: the magnetization significantly responds to the *magnitude*, but not the sign, of the polarization. As a result and as depicted in Fig. 5a with horizontal arrows, reversing the polarization from positive to negative between the two equilibrium Pmc2$_1$ phases– via, e.g., the application of an electric field along the pseudo-cubic [1-10] direction– should not change the direction nor the magnitude of the weak magnetization for both the $\Gamma_2$ and $\Gamma_4$ spin configurations. As shown in Fig. 5a too, this insensitivity to a polarization reversing of the weak magnetization in these two spin structures should also be found for the equilibrium Pmn2$_1$ phases – by, e.g., applying an electric field along [-1-10]. As schematized in Fig. 5b and indicated by Fig. 5a (by comparing, at the equilibrium polarization values, the solid and empty symbols having the same color), changing the magnetic structure from $\Gamma_2$ to $\Gamma_4$ within the equilibrium $Pmc2_1$ (or $Pmn2_1$) structure – via, e.g., the application of a magnetic field along [00-1] – has essentially no effect on the *magnitude* of the weak magnetization. In that case, the magnetization "only" rotates from the [110] to [00-1] pseudo-cubic directions. Furthermore, as also shown in Fig. 5a, the aforementioned response of the magnetization to polarization leads to a critical value of about 36 $\mu C/cm^2$ at which a full *vanishing* of the magnetization in the $Pmn2_1$ phase is obtained. In other words, we predict that the magnetization of the equilibrium $Pmn2_1$ phase can be greatly controlled when applying an electric field along the direction of its polarization. Such magnetization can even be reversed depending on the magnitude of this electric field.

Interestingly, there is yet another magneto-electric feature that is revealed by Fig. 5a, and which may also be put in use to design novel devices. Note that, for both the $\Gamma_2$ and $\Gamma_4$ magnetic configurations, the magnitude of the weak magnetization decreases faster with the magnitude of the polarization in the $Pmn2_1$ phase than it does in the $Pmc2_1$ phase. As a result, and as further stressed in Fig. 5a by means of oblique arrows, the structural path of Fig. 4 going from the $Pmn2_1$ ground state to the meta-stable $Pmc2_1$ state should result in a significant change of the magnitude of the weak magnetization. This change is about 0.038 $\mu_B$/f.u. for the $\Gamma_2$ spin configuration and about 0.035 $\mu_B$/f.u. for $\Gamma_4$, which correspond to an enhancement of the magnetization by almost of a factor of 2. In other words, as sketched in Fig. 5b, an electric-field-driven transition between the ground state and the metastable phase should be accompanied by a large magnetization change. This novel magneto-electric effect may also offer the opportunity to manipulate not only the magnetization amplitude but also its dynamics by applying successively electrical pulses in the [110] and [-110] pseudo-cubic directions, allowing to go back and forth between the ground state and the metastable phase. One can also envision to apply a dc electric field in one direction (e.g., [110]) and a pulse in the perpendicular direction (e.g., [-110]) to control with time the change of the magnetization amplitude.

**B) Development of a phenomenological model**

In order to explain all the results summarized in Fig. 5, we further develop a Landau-type phenomenological model with the energy given by:

$$
\begin{aligned}
E = &\kappa F^2 + \lambda F^2 P^2 + \alpha(\omega_{R,x} G_y F_z - \omega_{R,x} G_z F_y + \omega_{R,y} G_z F_x - \omega_{R,y} G_x F_z + \omega_{R,z} G_x F_y - \omega_{R,z} G_y F_x) \\
&+ \beta(P_x^2 \omega_{R,x} G_y F_z - P_y^2 \omega_{R,y} G_x F_z + P_z^2 \omega_{R,z} G_x F_y - P_z^2 \omega_{R,z} G_y F_x + P_y^2 \omega_{R,y} G_z F_x - P_x^2 \omega_{R,x} G_z F_y) \\
&+ \gamma_1 (P_x P_y \omega_{R,x} G_x F_z - P_x P_y \omega_{R,y} G_y F_z + P_z P_x \omega_{R,z} G_z F_y - P_z P_y \omega_{R,z} G_z F_x + P_y P_z \omega_{R,y} G_y F_x - P_z P_x \omega_{R,x} G_x F_y) \\
&+ \gamma_2 (P_x P_y \omega_{R,x} G_z F_x - P_x P_y \omega_{R,y} G_z F_y + P_z P_x \omega_{R,z} G_y F_z - P_z P_y \omega_{R,z} G_x F_z + P_y P_z \omega_{R,y} G_x F_y - P_z P_x \omega_{R,x} G_y F_z)
\end{aligned}
$$

(1)

where *x*, *y* and *z* subscripts refer to Cartesian components along the pseudo-cubic [100], [010] and [001] directions. **F** and **G** are the weak ferromagnetic vector and dominant G-type antiferromagnetic vector, respectively, while **P** is the electric polarization. $\omega_R$ is the vector characterizing the anti-phase oxygen octahedral tilting [29]: its direction is the axis about which the tilt occurs (i.e., the pseudo-cubic [-110] direction) and its magnitude is the rotation amplitude. Furthermore, *κ*, *λ*, *α*, *β*, $\gamma_1$ and $\gamma_2$ are material-dependent coefficients to be determined. The first term of Eq. (1) represents the usual harmonic effect associated with the magnetization, while the second energy characterizes the typical bi-quadratic interaction between magnetization and polarization [1]. The third term arises from the application of the universal law of Ref. [16] to a perovskite system possessing a dominant G-type antiferromagnetic vector altogether with anti-phase oxygen octahedral tilting (which leads to the occurrence of a weak magnetization). The last three energies of Eq. (1) have never been considered before, to the best of our knowledge, even if they are allowed by symmetry (in particular, they obey inversion- and time-reversal symmetry, and the sum of the k-points associated with all these terms adds up to zero). They characterize *penta-linear* interactions between different Cartesian components of the electrical polarization, anti-phase oxygen octahedral tiltings, G-type antiferromagnetic vector and magnetization.

Applying Eq. (1) to the four different cases depicted in Fig. 5 leads to the following

energies:

$$E(Pmn2_1, \Gamma_2) = \kappa F^2 + \lambda F^2 P^2 + \alpha GF\omega + \frac{\beta}{2} P^2 GF\omega - \frac{\gamma_2}{2} P^2 GF\omega$$

$$E(Pmn2_1, \Gamma_4) = \kappa F^2 + \lambda F^2 P^2 - \alpha GF\omega - \frac{\beta}{2} P^2 GF\omega - \frac{\gamma_1}{2} P^2 GF\omega$$

$$E(Pmc2_1, \Gamma_2) = \kappa F^2 + \lambda F^2 P^2 + \alpha GF\omega + \frac{\beta}{2} P^2 GF\omega + \frac{\gamma_2}{2} P^2 GF\omega$$

$$E(Pmc2_1, \Gamma_4) = \kappa F^2 + \lambda F^2 P^2 - \alpha GF\omega - \frac{\beta}{2} P^2 GF\omega + \frac{\gamma_1}{2} P^2 GF\omega$$

(2)

Here, F, ω, G and P are the projections of the magnetization, anti-phase tilting vector, antiferromagnetic vector and electrical polarization along their corresponding direction, respectively (e.g., F is the projection of the magnetization along the pseudo-cubic [110] direction for the $\Gamma_2$ magnetic configuration). Minimizing these energies with respect to the magnetization, and assuming that $\lambda P^2/\kappa$ is small with respect to unity, yields:

$$M(Pmn2_1, \Gamma_2) = -\frac{\alpha\omega G}{2\kappa}(1-\lambda P^2/\kappa) - \frac{(\beta-\gamma_2)P^2\omega G(1-\lambda P^2/\kappa)}{4\kappa}$$

$$M(Pmn2_1, \Gamma_4) = \frac{\alpha\omega G}{2\kappa}(1-\lambda P^2/\kappa) + \frac{(\beta+\gamma_1)P^2\omega G(1-\lambda P^2/\kappa)}{4\kappa}$$

$$M(Pmc2_1, \Gamma_2) = -\frac{\alpha\omega G}{2\kappa}(1-\lambda P^2/\kappa) - \frac{(\beta+\gamma_2)P^2\omega G(1-\lambda P^2/\kappa)}{4\kappa}$$

$$M(Pmc2_1, \Gamma_4) = \frac{\alpha\omega G}{2\kappa}(1-\lambda P^2/\kappa) + \frac{(\beta-\gamma_1)P^2\omega G(1-\lambda P^2/\kappa)}{4\kappa}$$

(3)

Remarkably, and as shown in Fig. 5 by means of solid and dashed lines, the first-principles results for the magnetization can be fitted extremely well by these Equations, the resulting fitting parameters being $\lambda/\kappa$=-2.85 m$^4$C$^{-2}$, $\alpha/\kappa$=-0.008 deg$^{-1}$, $\beta/\kappa$ =0.08 deg$^{-1}$m$^4$C$^{-2}$, $\gamma_1/\kappa$=0.04 deg$^{-1}$m$^4$C$^{-2}$, and $\gamma_2/\kappa$ =-0.04 deg$^{-1}$m$^4$C$^{-2}$. (These fits are done by taking $\omega_R$ and G to be constant and equal to 9.15 degrees and 2.39 $\mu_B$/f.u., respectively, as given by our LSDA results.) Such an excellent fit therefore attests the validity of the proposed Landau-type-model, which can thus be safely used to understand the numerical results depicted in Fig. 5. For instance, the right-hand side of Eqs. (3) only contain *even* orders (namely, second and fourth orders) of **P**, which explains why the magnetization depends on the magnitude, but not sign, of the polarization. The first term on the right-hand side of Eqs. (3) is also consistent with the first-principles results of Fig. 5a showing that the magnetization decreases in magnitude for increasing polarization values, since $\alpha/\kappa$ and $\lambda/\kappa$ have the same sign. Similarly, the second term on the right-hand side of Eqs. (3) also explains why the magnitude of the magnetization is more strongly reduced when increasing the polarization from zero in the $Pmn2_1$ state than in the $Pmc2_1$ phase for the $\Gamma_2$ (respectively, $\Gamma_4$) spin configuration, as $\beta/\kappa$ is positive while $\gamma_2/\kappa$ is negative (respectively, $\beta/\kappa$ and $\gamma_1/\kappa$ are both positive). As a result, the penta-linear couplings of Eq. (1), which lead to the $4^{th}$-order couplings of Eq. (3), are primordial to understand the unusual magneto-electric effects displayed by $CaMnO_3$ films.

### IV. Further Discussion

Importantly, these penta-linear terms should also be valid in other perovskites since their existence solely arises from symmetry considerations. In fact, these terms become active in structures derived from the orthorhombic phases that are most abundant among perovskite oxides, suggesting there are plenty of opportunities to find alternative materials presenting the effects described here, including at more moderate values of epitaxial strain. For example, $CaTiO_3$ is predicted to display the two polar phases of interest here at a moderate 2% epitaxial expansion [21]. This observation led us to consider $(CaTiO_3)_3/(CaMnO_3)_3$ superlattices in an attempt to reduce the epitaxial mismatched required to obtain the effects of interest. We found that, in such superlattices, our two polar states can be stabilized at an epitaxial mismatch of ~3.0%, therefore reducing the value of 3.9% obtained for pure $CaMnO_3$.

Our results also clearly indicate which are the experimental fingerprints of the presence and importance of such penta-linear couplings, i.e., large changes of magnetization when switching between states with differently-oriented polarization and/or a strongly anharmonic magnetoelectric response of a particular polar state. Further, Eq. (3) also implies the existence of other features, such as the magnetization reverting its direction when switching the anti-phase oxygen octahedral tilting or the G-type antiferromagnetic vector. Subsequent first-principles calculations confirm such features, further demonstrating the validity of our phenomenological theory. We thus hope that the penta-linear couplings revealed in this work will lead to the experimental discovery of novel magneto-electric materials, and that the variety of novel static and dynamical effects they permit will motivate the design of the next-generation of spintronic and magnonic devices.

**Acknowledgements**

This work was supported by the National Natural Science Foundation of China under Grants Nos. 51332006 and 11274270 (X.M.C.), the U.S. Department of Energy, Office of Basic Energy Sciences, under contract ER-46612 (L.B.), the ERC Consolidator Grant #615759 MINT (M.B.), FNR Luxembourg Grants FNR/P12/4853155/Kreisel (J.I.) and INTER/MOBILITY/15/9890527 GREENOX (L.B. and J.I.), and MINECO-Spain Grant MAT2013- 40581-P (J.I.).

**Figures and captions**

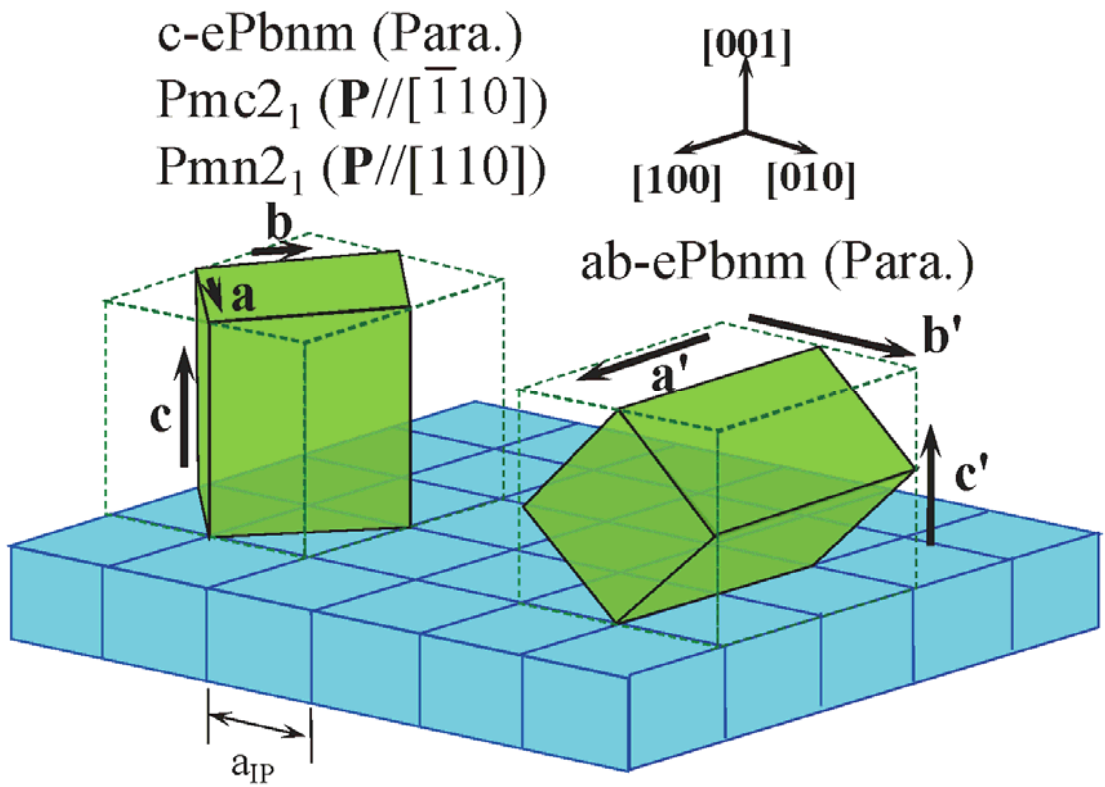

Fig. 1 (color online). Schematization of the growth of the different, presently considered structural phases of epitaxial CaMnO$_3$ thin films on cubic substrates.

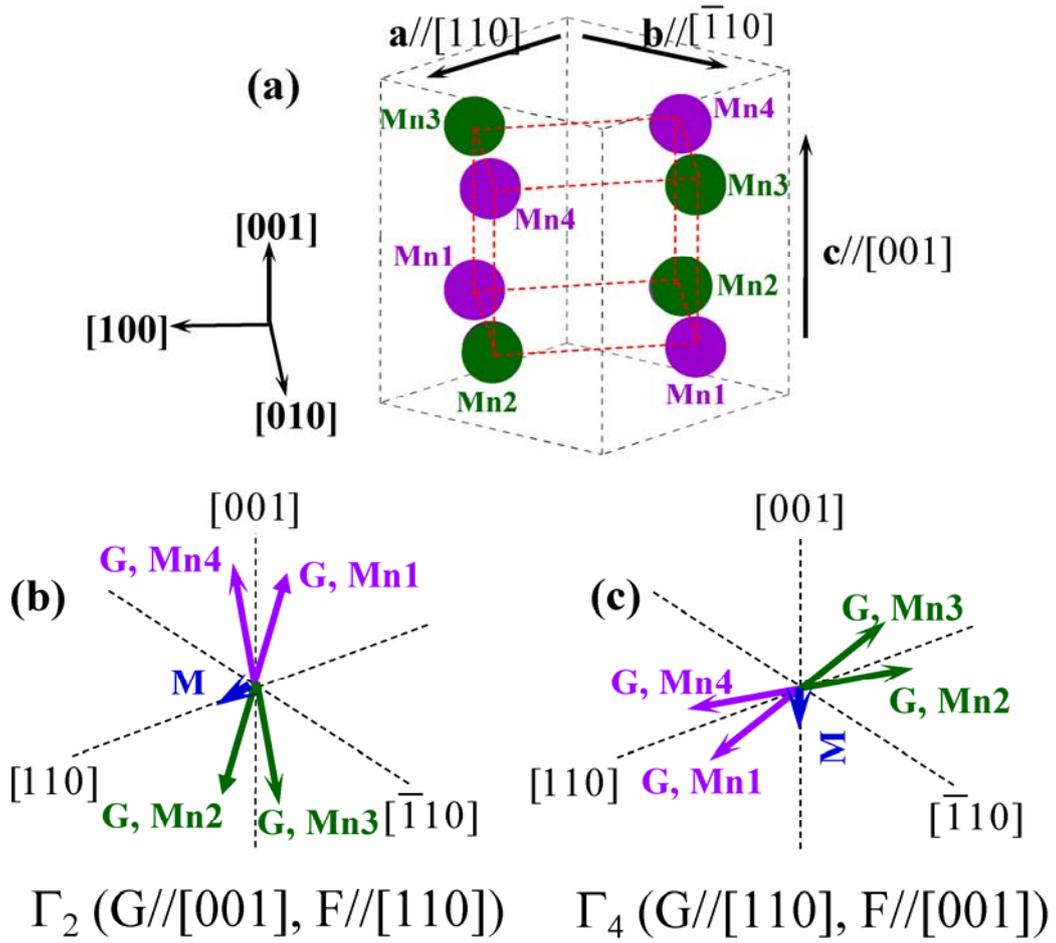

Fig. 2 (color online). Studied magnetic configurations in epitaxial CaMnO$_3$ thin films. Panel (a) depicts the different types of Mn atoms, while Panels (b) and (c) schematize the $\Gamma_2$ and $\Gamma_4$ spin configurations, respectively.

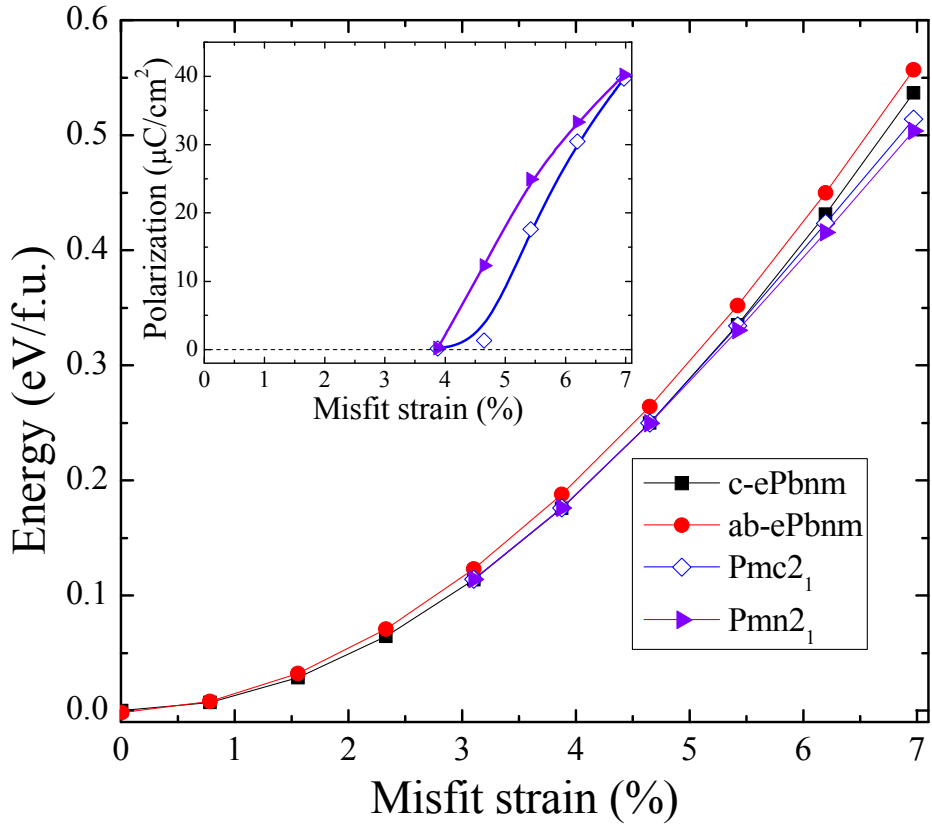

Fig. 3 (color online). Calculated total energy of epitaxial CaMnO$_3$ thin films (adopting a collinear G-type antiferromagnetic state) as a function of the tensile misfit strain, for the four considered structural phases. The inset shows the magnitude of the electric polarization versus misfit strain for the Pmc2$_1$ and Pmn2$_1$ states. The zero of energy corresponds to the c-ePbnm phase at zero strain.

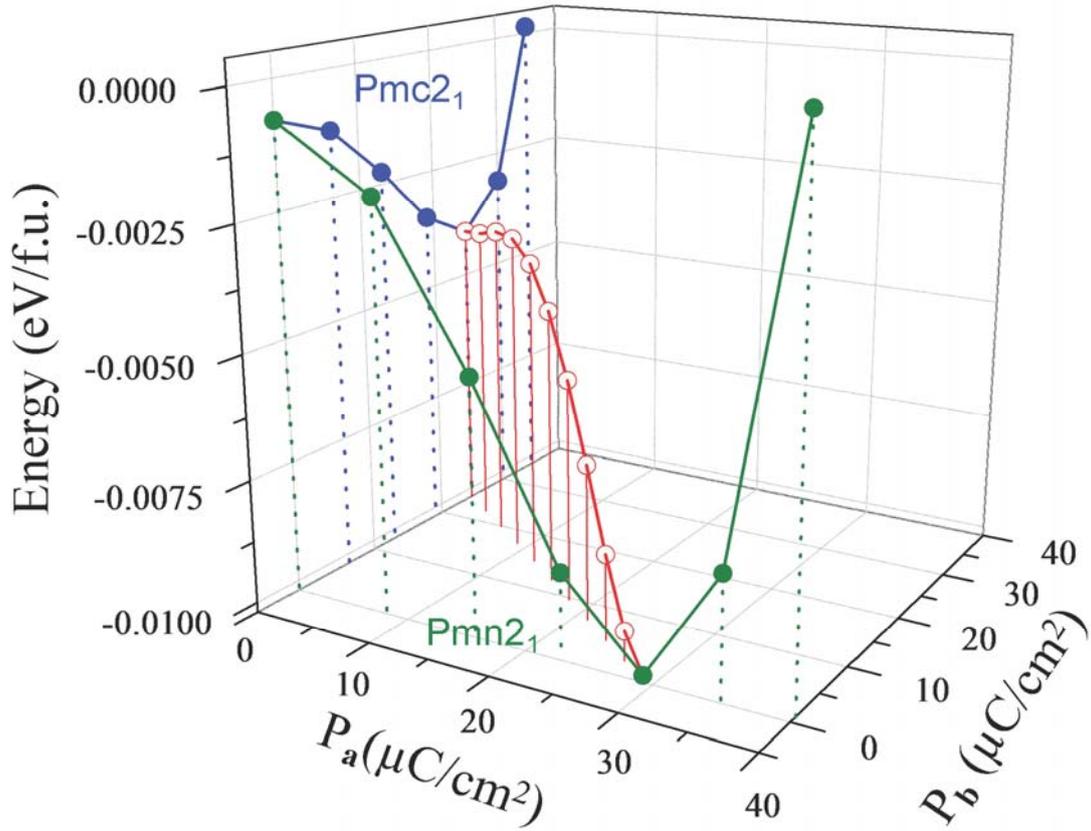

Fig. 4 (color online). Energy-*versus*-polarization curve for the Pmn2$_1$ (green line and symbols) and Pmc2$_1$ (blue line and symbols) phases of epitaxial CaMnO$_3$ thin films for a misfit strain of 5.8%. The red line and symbol show the energy of the intermediate structures connecting the Pmc2$_1$ and Pmn2$_1$ states, as a function of their electric polarization. P$_a$ and P$_b$ represent components of the polarization along the orthorhombic a and b axes, that are along the pseudo-cubic [110] and [-110] directions, respectively. The zero of energy is arbitrarily chosen to correspond to the Pmc2$_1$ state having a polarization of ~35 μC/cm$^2$ in magnitude for the misfit strain of 5.8%

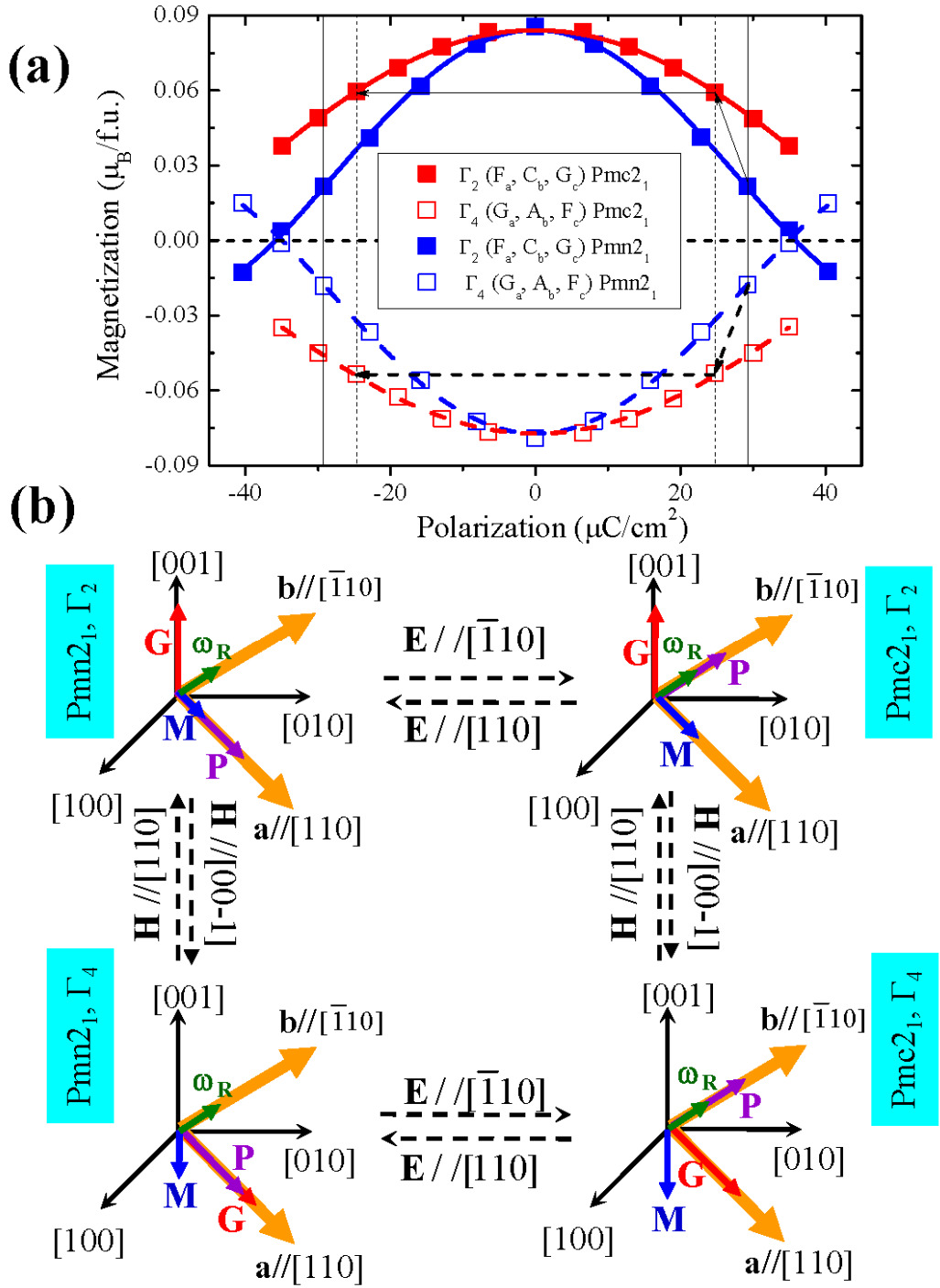

Fig. 5 (color online). Predicted magnetoelectric effects in epitaxial $CaMnO_3$ thin films being under a misfit strain of 5.8%. Panel (a) reports the magnetization-*versus*-polarization for the four different cases corresponding to the combination of the crystal structure among $Pmn2_1$ and $Pmc2_1$ *and* the choice of the magnetic structures among $\Gamma_2$ and $\Gamma_4$. This magnetization is along the pseudo-cubic [110] and [001] directions for the $\Gamma_2$ and $\Gamma_4$ spin configurations, respectively, and the polarization is along the pseudo-cubic [110] and [-110] directions for Pmn2$_1$ and Pmc2$_1$, respectively. The symbols represent the predictions from first-principles calculations, while the blue and red (solid and dashed) curves correspond to the fitting of these data by the Landau-type-derived Equation (3). The solid and dashed vertical lines indicate the value of the polarization in the equilibrium Pmn2$_1$ and Pmc2$_1$ states, respectively, while the horizontal dashed line depicts the zero in magnetization. The horizontal (respectively, oblique) solid arrow emphasizes the starting and ending magnetization, when the system undergoes a transition between two different but symmetry-equivalent Pmn2$_1$ minima having opposite polarization (respectively, between the equilibrium Pmn2$_1$ and Pmc2$_1$ states), in case of the $\Gamma_2$ spin configuration. Horizontal and oblique dashed arrows show similar data, but for the $\Gamma_4$ spin configuration. Panel (b) schematizes the different possible electric-field or magnetic-field driven transitions discussed in the text, and their effects on the directions and magnitude of the magnetization.